\documentclass[aps,prl,twocolumn,showpacs,floatfix]{revtex4} 
\usepackage{amsmath,bm,epsfig}

\begin{document}

\title{Anomalous Scaling and Refined Similarity of an Active Scalar in 
a Model of Homogeneous Turbulent Convection}
\author{Emily S. C. Ching$^{1,2}$ and W.C. Cheng$^1$}
\affiliation{$^1$ Department of Physics,
The Chinese University of Hong Kong, Shatin, Hong Kong}
\affiliation{$^2$ 
Institute of Theoretical Physics,
The Chinese University of Hong Kong, Shatin, Hong Kong}

\date{\today}
 
\begin{abstract}
Anomalous scaling in the statistics of an active scalar 
in homogeneous turbulent convection is studied using a 
dynamical shell model.
We extend refined similarity ideas
for homogeneous and isotropic turbulence to 
homogeneous turbulent convection and attribute the origin of the 
anomalous scaling to variations of the entropy transfer rate. 
We verify the consequences and thus the validity of our hypothesis 
by showing that the conditional statistics 
of the active scalar and the velocity at fixed values of entropy transfer rate 
are not anomalous but have simple scaling with exponents given by dimensional
considerations, and that the intermittency corrections are given
by the scaling exponents of the moments of the entropy transfer rate. 
\end{abstract}
\pacs{47.27.-i, 47.27.eb}
\maketitle

Since the work of Kolmogorov in 1941~(K41)~\cite{K41},
much effort has been devoted to the study of the possible universal statistics 
of fluid turbulence in the inertial range, the range of length scales that are smaller than
those of energy input and larger than those directly affected by molecular dissipation. 
A major challenge is to understand, 
from first principles, the origin of anomalous scaling, which is 
the deviation of the velocity scaling behavior from those predicted by 
dimensional considerations in K41. One important idea 
proposed by Kolmogorov in his refined theory~\cite{RSH}, which
we refer to as Kolmogorov's refined similarity idea, 
attributes the origin of anomalous scaling of the velocity to the variations of local 
energy dissipation rate. Kraichnan~\cite{Kraichnan} later pointed out that the local 
energy dissipation rate is not an inertial-range quantity and proposed
to attribute the origin of anomalous scaling of the velocity instead to 
the variations of the local energy transfer rate and we shall refer this as 
Kraichnan's refined similarity idea.

Similar problems of anomalous scaling can be posed for a scalar field 
advected by a turbulent velocity field. 
A passive scalar leaves the velocity statistics intact while an
active scalar couples with the velocity 
and influences its statistics. The problem of 
anomalous scaling of passive scalars is linear and has been recently
understood in terms of statistically preserved structures~\cite{FGVRMP2001}. 
On the other hand, the nonlinear problem of 
anomalous scaling of active scalars, like that of velocity, remains unsolved.  
A common example of an active scalar is temperature in turbulent convection
in which temperature variations drive the flow. 
Turbulent convection is often investigated experimentally 
in Rayleigh-B{\'e}nard convection cells heated from below and cooled on 
top~(see, e.g.,~\cite{Siggia,GL,Kadanoff} for a review). Such confined convective flows
are highly inhomogeneous with thermal and viscous boundary layers
near the top and the bottom of the cell. Moreover, coherent structures,
known as plumes, could affect the 
scaling properties~\cite{Ching2007}. For the purpose of studying anomalous
scaling of an active scalar, it would thus be more desirable to 
study homogeneous turbulent convection and in the absence of coherent structures.

Homogeneous turbulent convection has been proposed~\cite{Orszag}
as a convective flow in a box, with periodic 
boundary conditions, driven by a constant temperature gradient along the 
vertical direction. In Boussinesq approximation~\cite{Landau},
the equations of motion read~\cite{Ultimate}:
\begin{eqnarray}
\frac{\partial {\vec u}}{\partial t}
+ {\vec u} \cdot {\vec \nabla} {\vec u}
&=& -{\vec \nabla} p + \nu \nabla^2 {\vec u} + \alpha g \theta {\hat z}
\label{convection}  \\
\frac{\partial \theta}{\partial t}
+ {\vec u} \cdot {\vec \nabla} \theta &=& \kappa \nabla^2 \theta + \beta u_z 
\label{Teqn} 
\end{eqnarray}
with ${\vec \nabla} \cdot {\vec u} = 0$.
Here, ${\vec u}$ is the velocity, $p$ is
the pressure divided by the density,
$\theta = T- (T_0 - \beta z)$ is the deviation of temperature $T$ from 
a linear profile of constant temperature gradient of $-\beta$, 
$T_0$, $\alpha$, $\nu$, and $\kappa$ are respectively the mean temperature, the
volume expansion coefficient, kinematic viscosity and thermal diffusivity of the fluid, 
$g$ is the acceleration due to gravity, and
${\hat z}$ is a unit vector in the vertical direction.
The Bolgiano length~\cite{B59}, given by $L_B = \epsilon^{5/4} \chi^{-3/4} (\alpha g)^{-3/2}$,
where $\epsilon$ and $\chi$ are respectively the average energy and
thermal dissipation rates, is an estimate of a length scale above which 
buoyant forces are dominant. 
Numerical studies~\cite{Orszag,Biferale} revealed that
$L_B$ is of the order of the size of the periodic box. As a result,
temperature is not active in the intermediate scales. Indeed the small-scale
isotropic fluctuations were found~\cite{Biferale} to have scaling close 
to that of K41.

On the other hand, a dynamical shell model for homogeneous turbulent convection 
driven by a temperature gradient has also been proposed~\cite{Brandenburg}.
Shell model is constructed in a discretized Fourier space with 
$k_n= k_0 h^n$, $n = 0, 1, \ldots, N-1$, being the wavenumber in the $n$th shell,
and $h$ and $k_0$ are customarily taken to be 2 and 1 respectively.
Shell models for homogeneous and isotropic turbulence have been studied extensively
and proved to be successful in reproducing the scaling properties observed in
experiments~\cite{Shell}. In this shell model for homogeneous turbulent convection,
real variables $u_n$ and $\theta_n$ representing 
the Fourier transforms of $\vec{u}$ and $\theta$ 
satisfy the evolution equations:
\begin{eqnarray}
\nonumber
\frac{d u_n}{d t} + \nu k_n^2 u_n 
&=& ak_n(u_{n-1}^2-hu_nu_{n+1}) \\
 &+& bk_n(u_nu_{n-1}-hu^2_{n+1}) + \alpha g \theta_n 
\label{un} \\
\nonumber
\frac{d \theta_n}{d t} + \kappa k_n^2 \theta_n 
&=& \tilde ak_n(u_{n-1}\theta_{n-1}-hu_n\theta_{n+1}) \\
&+& \tilde b k_n(u_n \theta_{n-1}-hu_{n+1}\theta_{n+1}) + \beta u_n \ 
\label{thetan}
\end{eqnarray}
where $a$, $b$, $\tilde a$, and $\tilde b$ are positive parameters.
It was reported that the scaling behavior 
depends on the ratio $b/a$~\cite{Brandenburg}:
close to Bolgiano-Obukhov (BO) scaling~\cite{B59,O59} ($u_n \sim k_n^{-3/5}$, $\theta_n \sim k_n^{-1/5}$)
for $b/a$ large and close to K41 scaling~\cite{K41} 
($u_n \sim k_n^{-1/3}$, $\theta_n \sim k_n^{-1/3}$) for $b/a$ 
smaller than about 0.4.  
In this Letter, we show that 
buoyant forces are important when $b/a$ is large
and insignificant when $b/a$ is small in this shell model 
of homogeneous turbulent
convection and focus on $b/a$ large to
study anomalous scaling of an active scalar.
We extend refined similarity ideas for
homogeneous and isotropic turbulence to homogeneous turbulent convection.
Specifically, we extend Kraichnan's refined similarity idea
and attribute the origin of the anomalous scaling in homogeneous turbulent
convection to variations of the entropy transfer rate.
Using numerical simulations of the model, we verify the consequences and 
thus the validity of our hypothesis.

\begin{figure}[hbt]
\centering
\includegraphics[width=.45\textwidth]{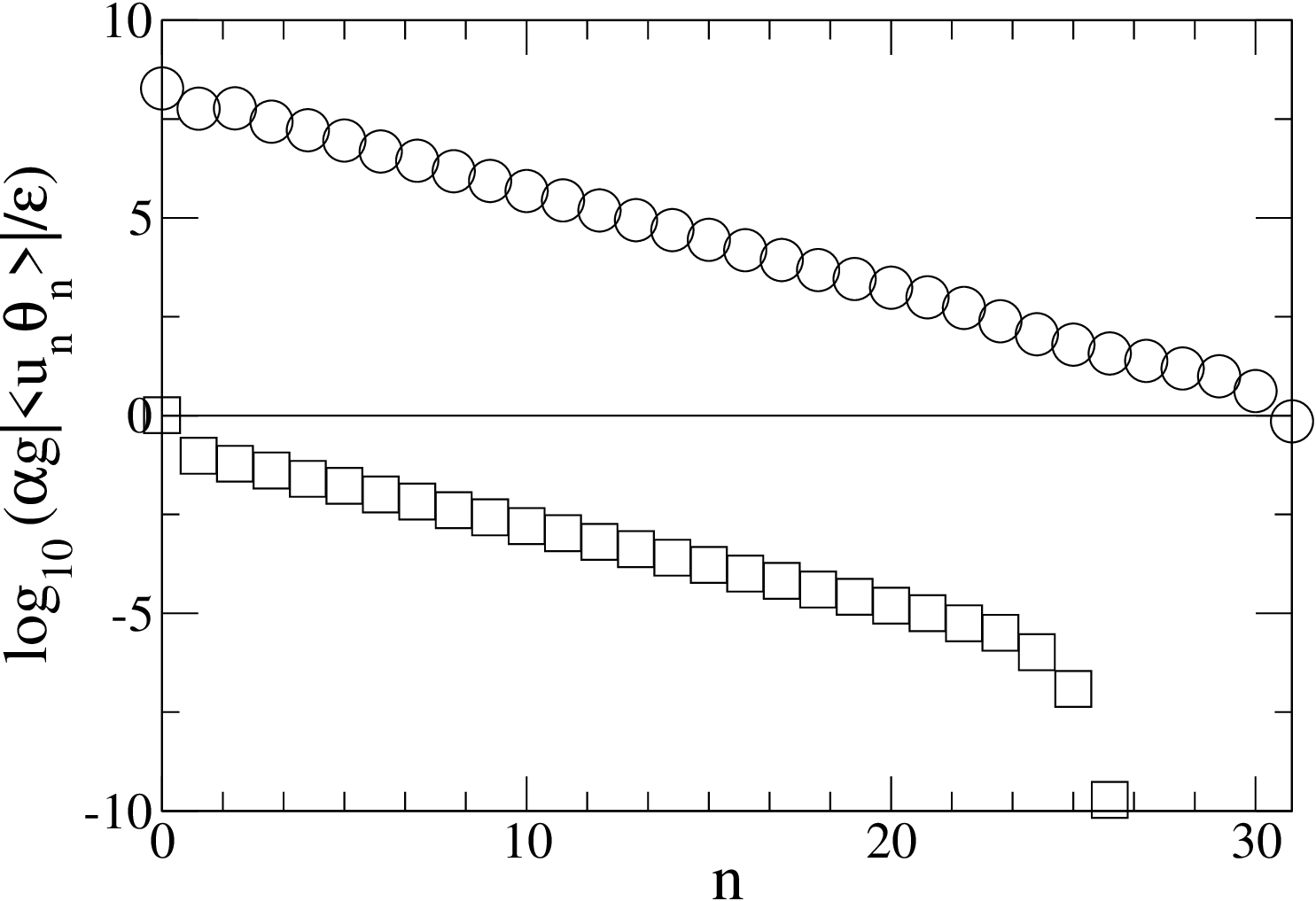}
\caption{The logarithm of $\alpha g |\langle u_n \theta_n\rangle|/\epsilon$
for different shells for $a=0.01, b=1$, $\beta = 1$, $\nu = 5 \times 10^{-17}$, 
$\kappa = 5 \times 10^{-15}$ and $N=32$~(circles), and
$a=10, b=1$, $\beta = 100$, $\nu=\kappa=10^{-8}$, and $N=30$~(squares) (the datapoints 
for $n= 28$ and $29$ are not shown here as they are too small.
For both cases, $\tilde a=\tilde b=1$ and $\alpha g=1$. }
\label{fig1}
\end{figure}

Multiply Eq.~(\ref{un}) by $u_n$, we get the energy budget:
\begin{equation}
\frac{d E_n}{dt} = F_u(k_n)-F_u(k_{n+1})-\nu k_n^2 u_n^2 + \alpha g u_n \theta_n
\label{En}
\end{equation}
where $E_n = u_n^2/2$ is the energy in the $n$th shell, 
$F_u(k_n) \equiv k_n(au_{n-1}+bu_n)u_{n-1}u_n$
is the rate of energy transfer from $(n-1)$th to $n$th shell,
$\nu k_n^2 u_n^2$ is the rate of energy dissipation in the 
$n$th shell due to viscosity, and $\alpha g u_n \theta_n$
is the power injected into the $n$th shell by the buoyant forces.
It is thus reasonable to take buoyancy to be significant in the $n$th shell if
\begin{equation}
\alpha g |\langle u_n \theta_n \rangle| > 
\epsilon \equiv \nu \sum_n k_n^2 \langle u_n^2 \rangle
\label{buoyant}
\end{equation}
where $\langle \ldots \rangle$ is an average over time. 
Note that for both K41 and BO scaling, 
$\alpha g \langle u_n \theta_n \rangle = \epsilon$ at $k_n = 1/L_B$.
We find that Eq.~(\ref{buoyant}) is satisfied for most of the shells and thus $\theta_n$ is
active only for $b/a$ is greater than about 2~(see Fig.~\ref{fig1}), 
and that this change coincides with the reported change in scaling 
behavior discussed above. 

\begin{figure}[hbt]
\centering
\includegraphics[width=.45\textwidth]{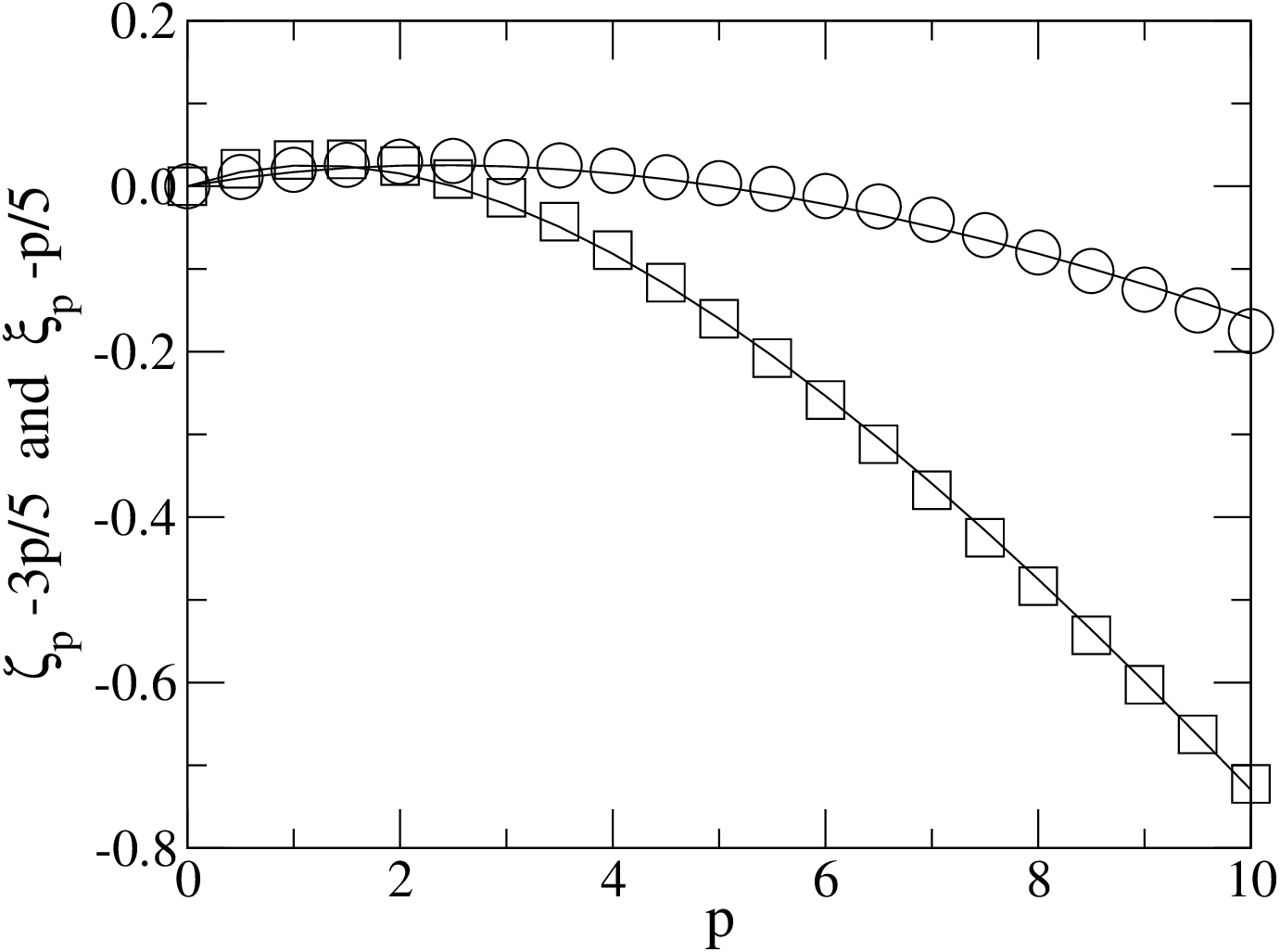}
\caption{Deviation of the scaling exponents from
the BO values:
$\zeta_p-3p/5$~(circles) and $\xi_p-p/5$~(squares).
The solid lines are the results of Eqs.~(\ref{anomalyzeta}) and (\ref{anomalyxi}).}
\label{fig2}
\end{figure}

As we are interested in the case of an active scalar, 
we focus on $b/a$ large and study 
the velocity and temperature structure functions,
$S_p(k_n)$ and $R_p(k_n)$:
\begin{equation}
S_p(k_n) \equiv \langle |u_n|^p \rangle \sim k_n^{-\zeta_p}  \ ; \ \ 
R_p(k_n) \equiv \langle |\theta_n|^p \rangle \sim k_n^{-\xi_p}
\label{SpRp}
\end{equation}
The scaling exponents $\zeta_p$ and $\xi_p$ 
do not depend on the values of the various parameters 
as long as $b/a$ is larger than 2. 
The results reported below are obtained using
$a=0.01, b=1$, $\beta = 1$, $\tilde a=\tilde b=1$, $\alpha g=1$,  $\nu = 5 \times 10^{-17}$,
$\kappa = 5 \times 10^{-15}$ and $N=32$.  As can be seen in Fig.~\ref{fig2}, 
both $\zeta_p$ and $\xi_p$ deviate respectively from 
the BO values of $3p/5$ and $p/5$ obtained from
dimensional considerations. Thus the active temperature and the velocity in homogeneous
turbulent convection have anomalous scaling.  
We study also the case where the temperature is driven
by a large-scale random forcing instead of an imposed linear gradient. 
In this case, the $\beta u_n$ term in Eq.~(\ref{thetan}) is
replaced by a random noise acting only in shell $n=0$. We find exactly the same scaling exponents,
supporting the universality of scaling of an active scalar upon different
forcing mechanisms~\cite{CMMV2002,ST95}. 

\vspace{0.5cm}

It was suggested~\cite{Lvov} that when buoyancy is dominant, the scaling behavior
of velocity and temperature spectra is governed by an entropy cascade of constant entropy flux.
In Bousinessq approximation, the entropy is proportional to the volume 
integral of temperature fluctuations. Entropy in the $n$th shell is, therefore,
defined as ${\cal S}_n \equiv \theta_n^2/2$. By studying the entropy 
budget obtained from
Eq.~(\ref{thetan}) upon mulitplication by $\theta_n$:
\begin{equation}
\frac{d {\cal S}_n}{dt} = F_\theta(k_n)-F_\theta(k_{n+1})-\kappa k_n^2 \theta_n^2 + 
\beta u_n \theta_n
\label{Sn}
\end{equation}
we get the rate of entropy transfer or entropy flux from $(n-1)$th to $n$th shell as:
\begin{equation}
F_\theta(k_n) \equiv k_n({\tilde a}u_{n-1}+
{\tilde b}u_n)\theta_{n-1}\theta_n
\end{equation}
In the stationary state and for intermediate scales where scaling is
observed, both 
$\beta \langle u_n \theta_n \rangle$ and
$\kappa k_n^2 \langle \theta_n^2 \rangle$ are negligible such that
there is indeed a constant entropy flux with $\langle F_\theta(k_n) \rangle
= \chi \equiv 
\kappa \sum_n k_n^2 \langle \theta_n^2 \rangle$.

\begin{figure}[thb]
\centering
\includegraphics[width=.45\textwidth]{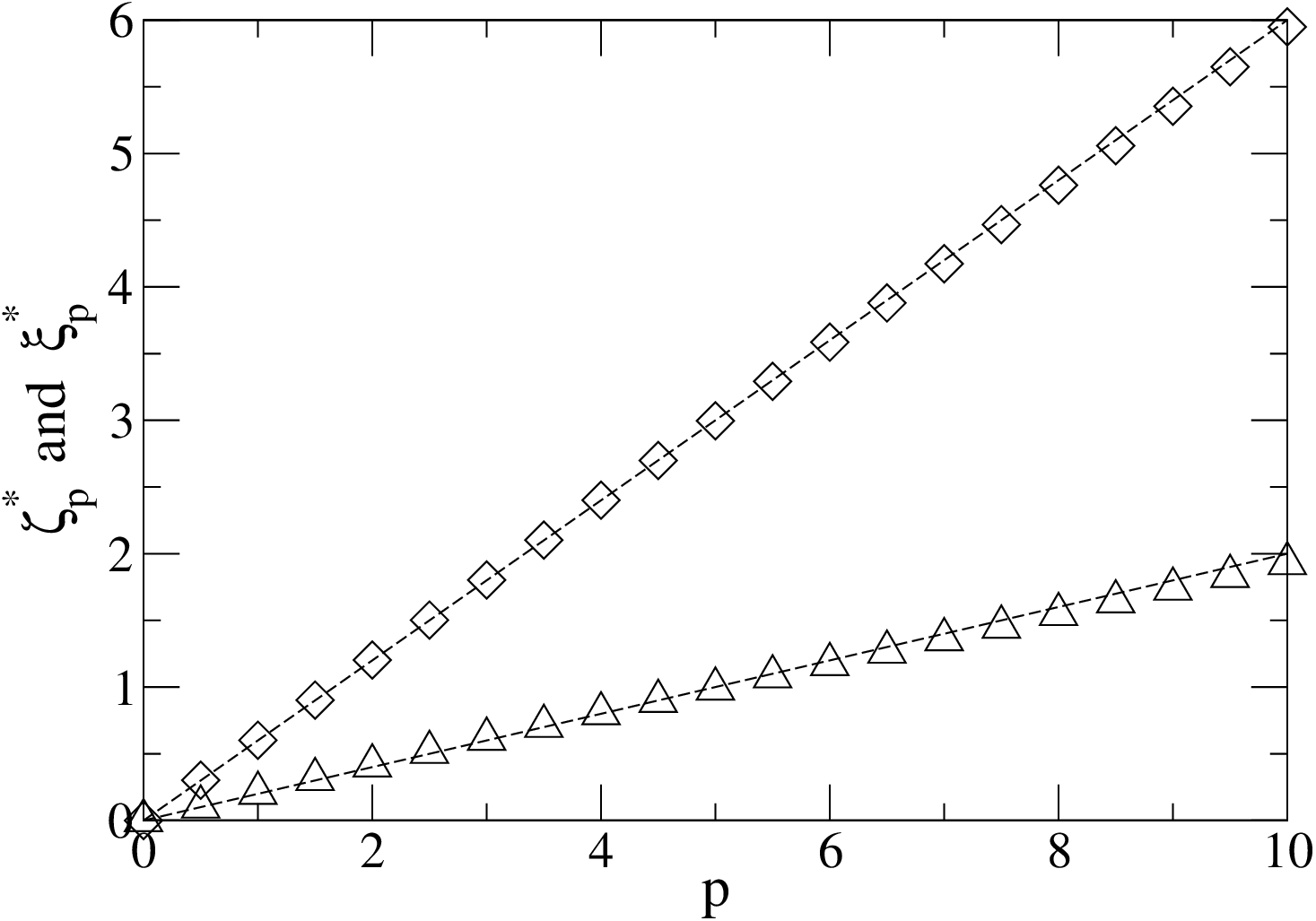}
\caption{The scaling exponents $\zeta_p^*$~(diamonds) and
$\xi_p^*$~(triangles)
of the conditional velocity and temperature structure functions. They are in 
good agreement with the BO values of $3p/5$ and $p/5$~(dashed lines).}
\label{fig3}
\end{figure}

We propose that when buoyancy is significant,
\begin{eqnarray}
u_n &=& \phi_u (\alpha g)^{2/5} |F_\theta(k_n)|^{1/5} k_n^{-3/5}
\label{RSHu}\\
\theta_n &=& \phi_\theta (\alpha g)^{-1/5} |F_\theta(k_n)|^{2/5} k_n^{-1/5}
\label{RSHtheta}
\end{eqnarray}
where $\phi_u$ and $\phi_\theta$ are dimensionless random variables 
that are independent of $k_n$
and statistically independent of $F_\theta(k_n)$. The absolute signs
are taken because the entropy flux $F_\theta(k_n)$, unlike $\chi$, 
can assume both positive and negative values. 
Equations.~(\ref{RSHu}) and
(\ref{RSHtheta}) are an extension of Kraichnan's refined similarity idea 
to homogeneous turbulent convection. 
With Eqs.~(\ref{RSHu}) and (\ref{RSHtheta}),
we attribute the anomalous scaling behavior of the active temperature and the velocity 
to the shell-to-shell variations of the entropy
transfer rate. An immediate consequence is 
that the conditional velocity and temperature structure functions at 
a certain prescribed value $x$ of the entropy transfer rate are given by
\begin{eqnarray}
\langle |u_n|^p \big| F_\theta(k_n)=x \rangle &=& 
\langle \phi_u^p \rangle (\alpha g)^\frac{2p}{5} x^\frac{p}{5}
k_n^{-\frac{3p}{5}} \sim k_n^{-\zeta_p^*}
\label{condu}\\
\langle |\theta_n|^p \big| F_\theta(k_n)=x \rangle &=&
\langle \phi_\theta^p \rangle (\alpha g)^{-\frac{p}{5}}
x^\frac{2p}{5} k_n^{-\frac{p}{5}} \sim k_n^{-\xi_p^*} \ , \ \
\label{condtheta}
\end{eqnarray}
and hence would have simple scaling 
with BO exponents of $\zeta_p^* = 3p/5$ and $\xi_p^* = p/5$
respectively. We evaluate the conditional velocity and temperature structure functions 
at different values of $x$ and confirm that $\zeta_p^*$ and $\xi_p^*$ are 
independent of $x$,
and in good agreement with 
$3p/5$ and $p/5$ respectively as shown in Fig.~\ref{fig3}.

\vspace{0.7cm}

\begin{figure}[thb]
\centering
\includegraphics[width=.45\textwidth]{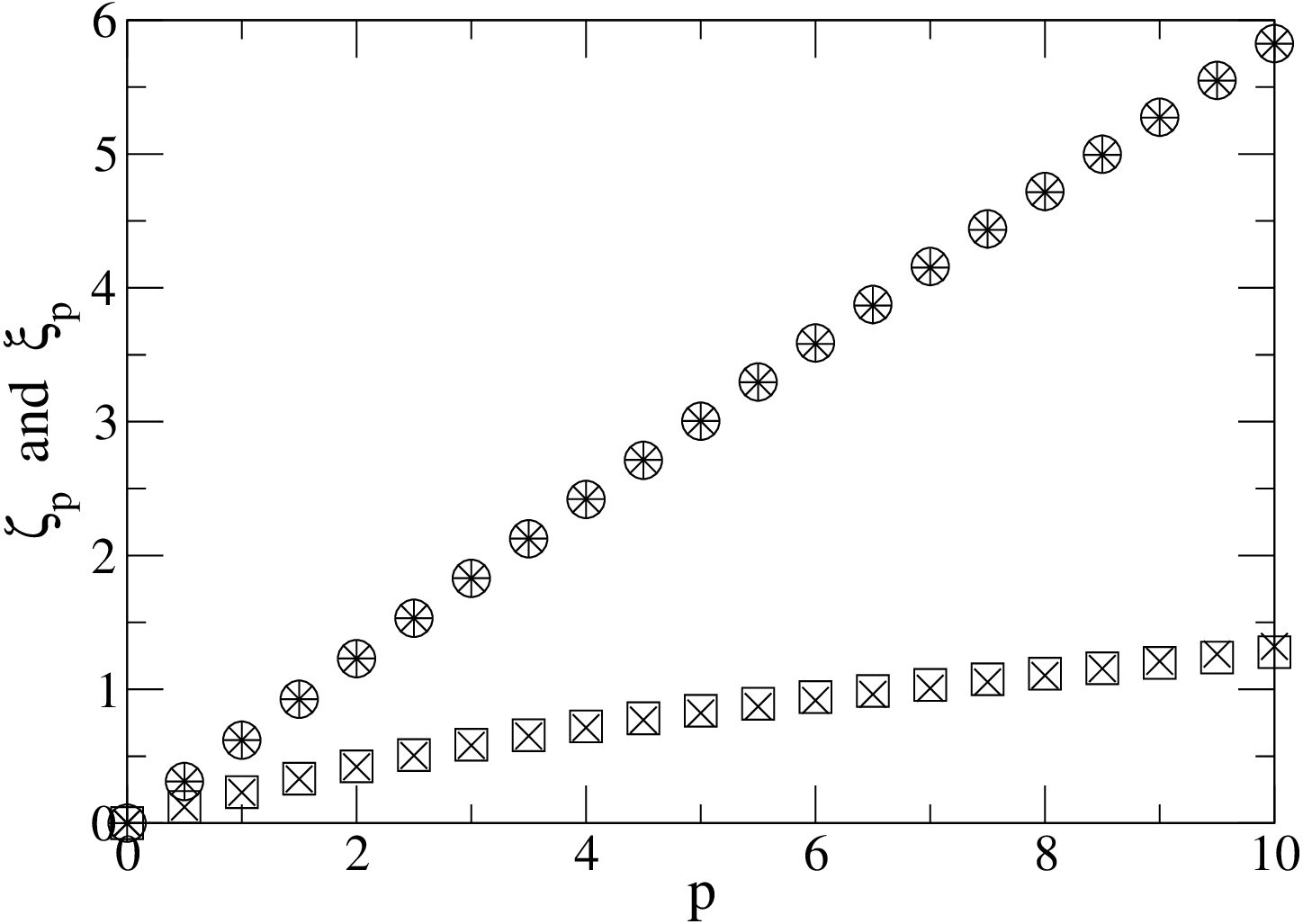}
\caption{Comparison of $\zeta_p$~(circles) and
$\xi_p$~(squares) with the theoretical predictions of $3p/5 + \tau_{p/5}$~(stars) and
$p/5+ \tau_{2p/5}$~(crosses) using the numerical results of $\tau_p$.}
\label{fig4}
\end{figure}

Let $\langle |F_\theta(k_n)|^p \rangle \sim k_n^{-\tau_p}$, then 
Eqs.~(\ref{RSHu}) and (\ref{RSHtheta}) imply
\begin{equation}
\zeta_p = {3 p}/{5} + \tau_{p/5} \ ; \qquad
\xi_p =  {p}/{5} + \tau_{2p/5}
\label{RSHzetaxi}
\end{equation}
showing that the intermittency corrections, which are the deviations of the scaling
exponents from the BO values, are given by the scaling exponents
of the moments of the entropy transfer rate.  
As the power of $F_\theta$ in $R_p$ is twice that in $S_p$, this explains
why the anomaly is larger for $\xi_p$ than for $\zeta_p$~(see Fig.~\ref{fig2}).
We evaluate $\tau_p$ numerically and check Eq.~(\ref{RSHzetaxi}) in Fig.~\ref{fig4}.
Good agreement is again found. 

\vspace{0.7cm}

Next, we show that the intermittency corrections, as given by $\tau_p$, 
can be obtained by suitably modifying the results of
the scaling exponents of the moments of the 
local thermal dissipation rate found in experiments~\cite{ChingKwok}.
In Ref.~\cite{ChingKwok},
the statistics of the local thermal dissipation rate, estimated by $\chi_\tau$,
have been studied in the central region of turbulent
Rayleigh-B{\'e}nard convection. It was found that the moments of 
$\chi_\tau \equiv (\langle u_c^2 \rangle \tau)^{-1} \int_t^{t+\tau} 
\kappa (\partial T/\partial t')^2 dt'$, where $\langle u_c^2 \rangle$ is the 
mean square velocity fluctuations at the centre, satisfy
a hierarchical structure of the She-Leveque form~\cite{SL}, 
and that their scaling exponents $\mu_p$, 
defined by $\langle \chi_\tau^p \rangle \sim \tau^{\mu_p}$, 
can be well described by $\mu_p = c(1-\beta_\chi^p)-\lambda p$ with $c = 1$, 
$\beta_\chi = 2/3$ and $\lambda=1/3$. The parameter $\lambda$ is
the scaling exponent of
$\lim_{p \to \infty} \langle \chi_\tau^{p+1} \rangle/
\langle \chi_\tau^{p} \rangle$, which was
estimated~\cite{ChingKwok}
as the ratio of the maximum thermal dissipation divided by a time $t_r$ 
at the scale $r=\sqrt{\langle u_c^2 \rangle}\tau$. Taking $t_r$ as $r/u_r$, $b=1/3$ implies
$u_r \sim r^{1/3}$, which is Kolmogorov scaling. 
Here, we find that the moments of the entropy transfer rate $\langle
|F_\theta(k_n)|^p \rangle$ also satisfy the same hierarchical structure~\cite{Cheng} 
and similarly $\tau_p$ is well approximated by 
 $c_1(1-\gamma^p)-c_2p$. Similarly, $c_2$ is the 
scaling exponent of $F_\theta^{(\infty)}(k_n) \equiv
\lim_{p \to \infty} 
\langle |F_\theta(k_n)|^{p+1} \rangle/\langle |F_\theta(k_n)|^p \rangle$.
Following Ref.~\cite{ChingKwok}, we estimate 
$F_\theta^{(\infty)}(k_n)$ as ${\cal S}_{max} u_n k_n$,
where ${\cal S}_{max}$ is the largest possible entropy. Since
we observe BO-like scaling in the present case,  
it is more appropriate to estimate $u_n \sim k_n^{-3/5}$. As a result, we
get $c_2 = 2/5$. Then as 
$\langle |F_\theta(k_n)| \rangle \sim \langle F_\theta(k_n)\rangle = \chi$ is
independent of $k_n$, we have $\tau_1 = 0$ implying $c_1(1-\gamma)=c_2$.
If we keep $c_1 = 1$ as $c=1$ for $\chi_\tau$,
then we get $\gamma = 3/5$. Putting these results together 
and using Eq.~(\ref{RSHzetaxi}), we find
\begin{eqnarray}
\zeta_p - {3p}/{5} &=& 1-({3}/{5})^{p/5}-{2}p/{25} 
\label{anomalyzeta} \\
\xi_p - {p}/{5} &=& 1-({3}/{5})^{2p/5}-{4}p/{25}
\label{anomalyxi}
\end{eqnarray}
Interesting, as shown in Fig.~\ref{fig2},  
Eqs.~(\ref{anomalyzeta}) and (\ref{anomalyxi}) indeed describe the measured intermittency
corrections well.

We have focussed on understanding the origin of anomalous scaling of an active scalar
in homogenous turbulent convection using a dynamical shell model. 
We have extended Kraichnan's refined similarity idea 
to an active scalar in homogeneous turbulent convection and attributed the anomalous scaling 
to the variations in the entropy transfer rate. 
We have verified our hypothesis by showing explicitly 
that the conditional velocity and temperature structure 
functions at fixed values of the entropy transfer rate have simple scaling exponents of 
the BO values, and the intermittency corrections are given by the scaling
exponents of the entropy transfer rate. Furthermore,
by modifying earlier results obtained for the statistics of the
local thermal dissipation rate in turbulent Rayleigh-B{\'e}nard
convection~\cite{ChingKwok}, we have obtained the scaling exponents $\tau_p$ of the moments
of the entropy transfer rate and thus Eqs.~(\ref{anomalyzeta}) and (\ref{anomalyxi})
for the intermittency corrections $\zeta_p - 3p/5$ and $\xi_p -p/5$. 
These results are found to be in good agreement with the numerical values obtained in the 
simulations of the shell model.

We should note that the scaling behavior of homogeneous turbulent convection might not be
the same as that in the central region of confined turbulent convection 
as coherent structures present in the latter case could affect the scaling
properties~\cite{Ching2007}. Indeed  
direct numerical simulations~\cite{Verzicco} and 
analyses of experimental data~\cite{JoT} 
indicated that the scaling behavior of the central region 
of confined turbulent convection is not well
described by BO scaling plus intermittency corrections. On the other hand,
there is evidence~\cite{ChingChau} of the validity of extending Kolmogorov's refined similarity idea
in terms of the local thermal dissipation rate in the central region of confined turbulent
convection. It would be interesting to further investigate this issue.

We thank TC Ko for his help in plotting Fig.~\ref{fig1}.
This work is supported in part by the Hong Kong Research Grants
Council (Grant No. CUHK 400304).


\begin{thebibliography}{99}
\bibitem{K41} A. N. Kolmogorov, Dokl. Akad. Nauk. SSSR {\bf 30}, 9
(1941), 
reproduced in Proc. R. Soc. London, Ser. A {\bf 434}, 9 (1991).
\bibitem{RSH} A. N. Kolmogorov, J. Fluid Mech. {\bf 12}, 82 (1962).
\bibitem{Kraichnan} R.H. Kraichnan, J. Fluid Mech. {\bf 62}, 305 (1974).
\bibitem{FGVRMP2001} G. Falkvoich,  K. Gawedzki, M. Vergossola, 
Rev. Mod. Phys. {\bf 73}, 913 (2001).
\bibitem{Siggia} E.D. Siggia, Ann. Rev. Fluid Mech. {\bf 26}, 137 (1994).
\bibitem{GL} S. Grossmann and D. Lohse, J. Fluid Mech. {\bf 407}, 27 (2000).
\bibitem{Kadanoff} L.P. Kadanoff, Phys. Today {\bf 54}(8), 34 (2001).
\bibitem{Ching2007} E.S.C. Ching, Phys. Rev. E {\bf 75}, 056302 (2007).
\bibitem{Orszag} V. Borue, S.A. Orszag, J. Sci. Comput. {\bf 12}, 305 (1995).
\bibitem{Landau} See, for example, Landau and Lifshitz, {\it Fluid
Mechanics} (Pergamon Press, Oxford, 1987).
\bibitem{Ultimate} D. Lohse and F. Toschi, Phys. Rev. Lett. {\bf 90}, 034502 (2003).
\bibitem{B59} R. Bolgiano, J. Geophys. Res. {\bf 64}, 2226 (1959).
\bibitem{Biferale} L. Biferale, E. Calzavarini, F. Toschi, R. Tripiccione, Europhys. Lett.
{\bf 64}, 461 (2003).
\bibitem{Brandenburg} A. Brandenburg, Phys. Rev. Lett. {\bf 69}, 605 (1992).
\bibitem{Shell} See, for example, L. Biferale, Annu. Rev. Fluid Mech. {\bf 35}, 441 (2003).
\bibitem{O59} A.M. Obukhov, Dpkl. Akad. Nauk. SSSR {\bf 125}, 1246 (1959).
\bibitem{CMMV2002} 
A. Celani, T. Matsumoto, A. Mazzino, and M. Vergassola, Phys. Rev. Lett. {\bf
88}, 054503 (2002).
\bibitem{ST95} E. Suzuki and S. Toh, Phys. Rev. E {\bf 51}, 5628 (1995).
\bibitem{Lvov} V.S. L'vov, Phys. Rev. Lett. {\bf 67}, 687 (1991).
\bibitem{ChingKwok} E.S.C. Ching and C.Y. Kwok, Phys. Rev. E {\bf 62}, R7587 (2000).
\bibitem{SL} Z.-S. She and E. Leveque, Phys. Rev. Lett. {\bf 72}, 336 (1994).
\bibitem{Cheng} W.C. Cheng, M.Phil. Thesis, The Chinese University of Hong Kong (2007).
\bibitem{Verzicco} R. Verzicco and R. Camussi, J. Fluid Mech. {\bf 477}, 19 (2003); R. Camussi and R.
Verzicco, Europhy. J. Mech. B {\bf 23}, 427 (2004).
\bibitem{JoT} E.S.C. Ching, K.W. Chui, X.-D. Shang, P. Tong and K.-Q. Xia, J. Turb. {\bf 5}, 27
(2004).
\bibitem{ChingChau} E.S.C. Ching and K.L. Chau, Phys. Rev. E {\bf 63}, 047303 (2001).
\end{thebibliography}
\end{document}